\documentclass[useAMS,usenatbib]{mnras}
\usepackage[pdftex]{graphicx} 
\usepackage{hyperref}
\usepackage{xcolor} 
\usepackage{lscape}
\usepackage{calc}
\usepackage{amsmath}
\usepackage{afterpage}
\usepackage{bm}
\usepackage{amssymb, physics}
\defcitealias{Jespersen2025a}{J25}

\hypersetup{colorlinks,linkcolor={red!50!black}, citecolor={blue!50!black},urlcolor={blue!80!black} } 

\title[Later evolution of protoclusters]
{On the later evolution of observationally selected protocluster candidates at $z\,{\gtrsim}\,5$}

\author[Lim et al.]
{Seunghwan Lim$^{1,2}$\thanks{E-mail: sl2207@cam.ac.uk}
\\
\vspace*{6pt} \\
$^{1}$Kavli Institute for Cosmology, University of Cambridge, Madingley
Road, Cambridge, CB3 0HA, UK \\
$^{2}$Cavendish Laboratory, University of Cambridge, 19 JJ Thomson
Avenue, Cambridge, CB3 0HE, UK 
\
}

\begin{document} 

\pagerange{\pageref{firstpage}--\pageref{lastpage}}

\date{\today}
\pubyear{2026}

\maketitle

\label{firstpage}

\begin{abstract}
Recent observations have revealed numerous protocluster candidates at $z\,{\gtrsim}\,5$, yet whether these systems will eventually evolve into today's galaxy clusters remains an open question. Using the FLAMINGO simulations---resolving protocluster cores up to $z\,{\simeq}\,10$---we track the later evolution of observationally selected protocluster candidates, comparing three selection methods against observational samples. The observed number density falls between our mass-selected and abundance-matched samples, implying that current searches pick up both genuine cluster progenitors and significant interlopers that will not reach cluster masses by $z\,{=}\,0$. We find that candidates at $z\gtrsim5$ are heavily clustered, hosting 2$-$10 neighbors within 10\,cMpc. Consequently, a candidate with a neighbor at 5\,cMpc (10\,cMpc) faces a $\gtrsim50\%$ ($\gtrsim30\%$) probability of later merging into a larger system, mostly at $z\,{\lesssim}\,2$. The merger count converges beyond ${\sim}10$\,cMpc, pointing to a fundamental scale in structure formation. Observations show markedly weaker clustering than our simulations predict, suggesting clustering offers a currently overlooked diagnostic. Each candidate undergoes roughly 2$-$6 later major mergers, mostly with systems too small to be recognized as massive at the selection epoch. Hence, relying solely on high-$z$ mass and galaxy overdensity to forecast a candidate's fate is prone to severe scatter and systematic error. A robust identification of true cluster progenitors demands a total mass sum of galaxies down to the faintest levels within a 10\,cMpc radius. Upcoming surveys with both depth and area will be key to reliably linking high-$z$ protocluster candidates to their ultimate destiny.
\end{abstract} 

\begin{keywords}
galaxies: evolution -- galaxies: high-redshift -- galaxies: clusters: general -- early Universe -- large-scale structure of Universe -- methods: statistical
\end{keywords}

\section[Introduction]{Introduction}
\label{sec_intro}

Protoclusters, the progenitors of today's galaxy clusters, represent the most overdense regions in the high-redshift Universe and serve as unique laboratories for studying galaxy formation and large-scale structure evolution under extreme environmental conditions \citep[e.g.,][]{Chiang2013, Muldrew2015, Casey2016, Overzier2016, Lim2021, Alberts2022}. Identifying protoclusters at $z\gtrsim5$ and tracking their evolution to the present day is essential for understanding how the most massive structures in the Universe assemble, how galaxy properties depend on environment at early times, and for testing the $\Lambda$CDM paradigm on the largest scales \citep[e.g.,][]{Chiang2017, Lovell2018, Lim2024, Sun2024, Witten2026}. 

Recent observational campaigns, particularly with the James Webb Space Telescope (JWST) and Subaru Telescope, have dramatically increased the number of known protocluster candidates at $z\gtrsim5$ \citep[e.g.,][]{Harikane2019, Higuchi2019, Laporte2022, Brinch2023, Morishita2023, Helton2024b, Champagne2025, Li2025, WuC2025, Terp2026, WuZ2026}. Importantly, there have been spectroscopic, systematic searches from large survey volumes that enable statistical samples of protocluster candidates at these redshifts for the first time \citep[e.g.,][]{Higuchi2019, Helton2024b, Terp2026}. \citet{Higuchi2019} identified 40 Lyman-$\alpha$ emitter-based candidates at $z\sim5.7$ and $6.6$ over wide-area Subaru SILVERRUSH fields. \citet{Helton2024b} reported seventeen candidates at $5\lesssim z\lesssim9$ combining JADES and JEMS imaging with FRESCO spectroscopy. More recently, \citet{Terp2026} discovered six overdensities at $z\sim5.5-7$ using JWST/NIRCam grism spectroscopy of the Abell 2744 lensing field. These discoveries have opened a new window into the epoch of reionization, yet they have also raised a critical question: are these observed overdensities truly destined to evolve into Coma- or Virgo-like clusters by $z=0$?

Despite the growing observational sample, connecting high-redshift protocluster candidates to their low-redshift descendants remains challenging. A fundamental issue lies in the inconsistency of definitions (see \citealt{Lim2021} for a detailed discussion) and apertures employed across different studies \citep{Lim2024, Witten2026}. Aperture sizes used to identify protoclusters range from the virial radius of the central halo (a few tens of physical kpc) to the entire Lagrangian region that will eventually collapse into a cluster (of order 10 comoving Mpc), spanning nearly two orders of magnitude. As demonstrated by \citet{Lim2024} and also confirmed by \citet{Witten2026}, this aperture variation can lead to order-of-magnitude differences in estimated total mass, stellar mass, and star formation rate for a protocluster candidate at $z\gtrsim4$. Furthermore, \citet{Witten2026} showed that true cluster progenitors constitute only a small fraction of all objects with similar mass at high redshift; the majority instead evolve into group-scale systems ($M_{\rm h}\sim10^{13}-10^{14}\,\mathrm{M}_{\odot}$) by $z=0$. Studies have also revealed a large scatter in the galaxy overdensity among cluster progenitors at $z\,{\gtrsim}\,5$, as well as a scatter in completeness and purity at a given overdensity threshold \citep[e.g.,][]{Chiang2013, Lovell2018, Morokuma-Matsui2025}. This suggests that selection based primarily on mass and galaxy overdensity, as commonly practiced, is prone to significant contamination and systematic bias.

An alternative and potentially powerful diagnostic is clustering. The clustering strength of protocluster candidates directly informs their host halo masses and merger histories \citep{Muldrew2015}. Moreover, the proximity of multiple massive systems at high redshift may indicate future mergers that can alter the evolutionary track of individual candidates. If protocluster candidates are strongly clustered among themselves, a substantial fraction may not survive as isolated progenitors but instead merge into even more massive systems by $z=0$. Conversely, if they are isolated, their eventual fate may be more reliably predicted from their current properties. Quantifying this clustering and its implications for later mergers is thus essential for interpreting the growing census of high-$z$ protocluster candidates.

In this study, we address these outstanding questions using the FLAMINGO suite of hydrodynamical simulations \citep{Schaye2023, Kugel2023}. FLAMINGO is among the largest-volume hydrodynamic simulations, with a flagship $(1\,\mathrm{cGpc})^3$ run that resolves protocluster cores up to $z\sim10$, making it ideally suited for studying rare, massive structures and also quantifying cosmic variance. We construct three different protocluster samples in the simulations---ranging from a fiducial mass-selected sample that approximates observational selections, to abundance-matched and true main progenitor samples---and compare their properties against the observational samples from systematic searches over large volumes with spectroscopic data. We then analyze the clustering of these simulated systems, compute merger probabilities as a function of neighbor distance, and quantify the number of later mergers each candidate experiences. Finally, we discuss the implications for observational protocluster searches and propose an optimal strategy for identifying true cluster progenitors at high redshift.

The structure of this paper is as follows. In Section \ref{sec_data}, we describe the FLAMINGO simulations and the observational samples used for comparison. Section \ref{sec_method} presents our three definitions of protocluster samples and our approach to evaluating cosmic variance. In Section \ref{sec_results}, we present our results on the abundance, total mass, clustering, and merger properties of protocluster candidates. Section \ref{sec_discussion} interprets these findings in the context of whether observed candidates are true cluster progenitors and what constitutes an optimal identification strategy. Finally, Section \ref{sec_summary} summarizes our main conclusions and outlines future prospects. 

\section[Data]{Data}
\label{sec_data}

\subsection{The FLAMINGO simulations}

Our analysis is based on the predictions from the FLAMINGO simulations \citep{Schaye2023, Kugel2023}. Because the number density of protoclusters is low, a large simulation box is essential for studying them, particularly for taking cosmic variance into account \citep{Lim2024, Lim2025b}. The FLAMINGO simulations are among the largest hydrodynamical simulations, with box sizes of order a comoving Gpc (cGpc), making them suitable for this purpose. The FLAMINGO data release is presented in \citet{Helly2026}. The simulations are run using the \textsc{Swift} numerical solver \citep{Schaller2024}, which employs smoothed-particle hydrodynamics (SPH) to solve the equations of hydrodynamics.

Subgrid physics prescriptions model processes occurring below the resolution limit. Radiative cooling and heating rates follow \citet{PloeckingerSchaye2020}, while the multi-phase interstellar medium (ISM) pressure and temperature are treated following \citet{SchayeDallaVecchia2008}. Star formation proceeds by converting gas particles with hydrogen number density $n_{\mathrm{H}} > 10^{-1}\,\mathrm{cm}^{-3}$ into stellar particles in a stochastic manner, designed to reproduce the Kennicutt-Schmidt law \citep{SchayeDallaVecchia2008}. The simulations assume a \citet{Chabrier2003} initial mass function (IMF). The subgrid models for stellar feedback, including supernovae and winds, follow the kinetic models of \citet{Chaikin2023}.

Black holes are seeded in haloes above a threshold mass (e.g., $M_{\mathrm{h}} > 2.8\times10^{10}\,\mathrm{M}_{\odot}$ for the highest-resolution run), with an initial subgrid mass of $10^{5}\,\mathrm{M}_{\odot}$. Black holes grow via a modified Bondi-Hoyle accretion rate \citep{Springel2005}, enhanced in dense regions \citep{BoothSchaye2009}, and are repositioned to the local potential minimum at each timestep to account for unresolved dynamical friction \citep{Bahe2022}. AGN feedback is implemented via a thermal model \citep{BoothSchaye2009} for the high-resolution runs, where feedback energy is injected stochastically into neighbouring gas particles once a sufficient energy threshold is reached. Free parameters in the subgrid models, particularly those governing stellar and AGN feedback, are calibrated using a machine learning approach to match the observed low-redshift stellar mass function and cluster gas fractions \citep{Kugel2023}.

Halo catalogues and merger trees are generated using \textsc{HBT-HERONS} \citep{ForouharMoreno2025}, an updated version of the history-based subhalo finder HBT+ \citep{Han2018}. \textsc{HBT-HERONS} improves upon HBT+ by enhancing subhalo tracking in hydrodynamical simulations and producing more robust merger trees with fewer catastrophic failures. A wide range of halo and galaxy properties are computed using the SOAP post-processing tool \citep{McGibbon2025}, which calculates quantities consistently across different subhalo finders. Our fiducial galaxy properties are integrated within a 3D aperture of 50 physical kpc (pkpc) from the galaxy centre. For haloes, properties including mass are defined as those within $R_{200}$, the 3D radius within which the mean density is 200 times the critical density.

The FLAMINGO suite includes simulations with a variety of box sizes, resolutions, subgrid models, and cosmologies. Among these, our analysis is based on their flagship high-resolution run, `L1\_m8', which simulates a $(1\,\mathrm{cGpc})^3$ comoving volume with an initial baryonic gas particle mass of $m_{\mathrm{gas}} = 1.34\times10^{8}\,\mathrm{M}_{\odot}$. This resolution resolves galaxies (haloes) of mass $\sim10^{10}\,\mathrm{M}_{\odot}$ ($\sim10^{11}\,\mathrm{M}_{\odot}$) with 100 particles. We choose this highest-resolution run because it is the only FLAMINGO run that resolves the cores of protoclusters up to $z\simeq10$. The simulation assumes the Dark Energy Survey Year 3 (DES Y3) cosmology \citep{Abbott2022} for a flat universe.

\subsection{Observational samples of high-$z$ protoclusters}

To compare with the predictions from FLAMINGO, we use the observational samples of protoclusters from \citet{Higuchi2019}, \citet{Helton2024b}, and \citet{Terp2026}. We select these three studies because they performed systematic, volume-limited searches for high-$z$ galaxy overdensities with spectroscopy. As the primary goal of our study is a census of protoclusters at high redshift and its implications for their evolution, consistent samples from uniform searches are required. Given that \citet{Helton2024b} and \citet{Terp2026} are based on JWST data, while \citet{Higuchi2019} used Subaru telescope data, this combination may reveal and mitigate any survey- or instrument-dependent systematics.

\citet{Higuchi2019} identified protocluster candidates from the SILVERRUSH survey \citep{Ouchi2018} using Lyman-$\alpha$ emitter (LAE) samples. They discovered 14 and 26 candidates at $z\simeq5.7$ and $z\simeq6.6$, respectively, over sky areas of $13.8\,\mathrm{deg}^2$ and $16.2\,\mathrm{deg}^2$. The effective comoving survey volumes are $1.1\times10^7\,\mathrm{cMpc}^3$ and $1.5\times10^7\,\mathrm{cMpc}^3$, yielding number densities of $1.2\times10^{-6}\,\mathrm{cMpc}^{-3}$ and $1.7\times10^{-6}\,\mathrm{cMpc}^{-3}$ for the two redshift bins. Halo masses of the candidates were estimated using the theoretical model of \citet{Inoue2018}, by matching the average mass of haloes that produce similar overdensities.

\citet{Helton2024b} identified seventeen protocluster candidates at $5\lesssim z\lesssim9$ by combining multiple JWST surveys. Specifically, candidates were pre-selected from JADES \citep{Eisenstein2023} and JEMS \citep{Williams2023}, and confirmed via FRESCO spectroscopy \citep{Oesch2023} using H$\alpha$ and [OIII]$\lambda5008$ detections, covering $4.9<z<6.6$ and $6.7<z<8.9$, respectively. The overlapping survey area is approximately $81\,\mathrm{arcmin}^2$, corresponding to comoving volumes of $\sim3.2\times10^5\,\mathrm{cMpc}^3$ (for H$\alpha$) and $\sim3.3\times10^5\,\mathrm{cMpc}^3$ (for [OIII]). Candidates were identified using a Friends-of-Friends (FoF) algorithm with fixed linking parameters: a projected distance of 500 pkpc and a line-of-sight velocity dispersion of 500 km s$^{-1}$, requiring the galaxy number density within the aperture to meet $N_{\rm gal}\geq4$ and $\delta_{\rm gal}\geq3$. Halo masses were then inferred using the empirical model of \citet{Behroozi2019}. \citet{Helton2024b} found that varying the linking parameters by a factor of a few does not strongly alter their results.

\begin{figure*}
\includegraphics[width=1.\linewidth]{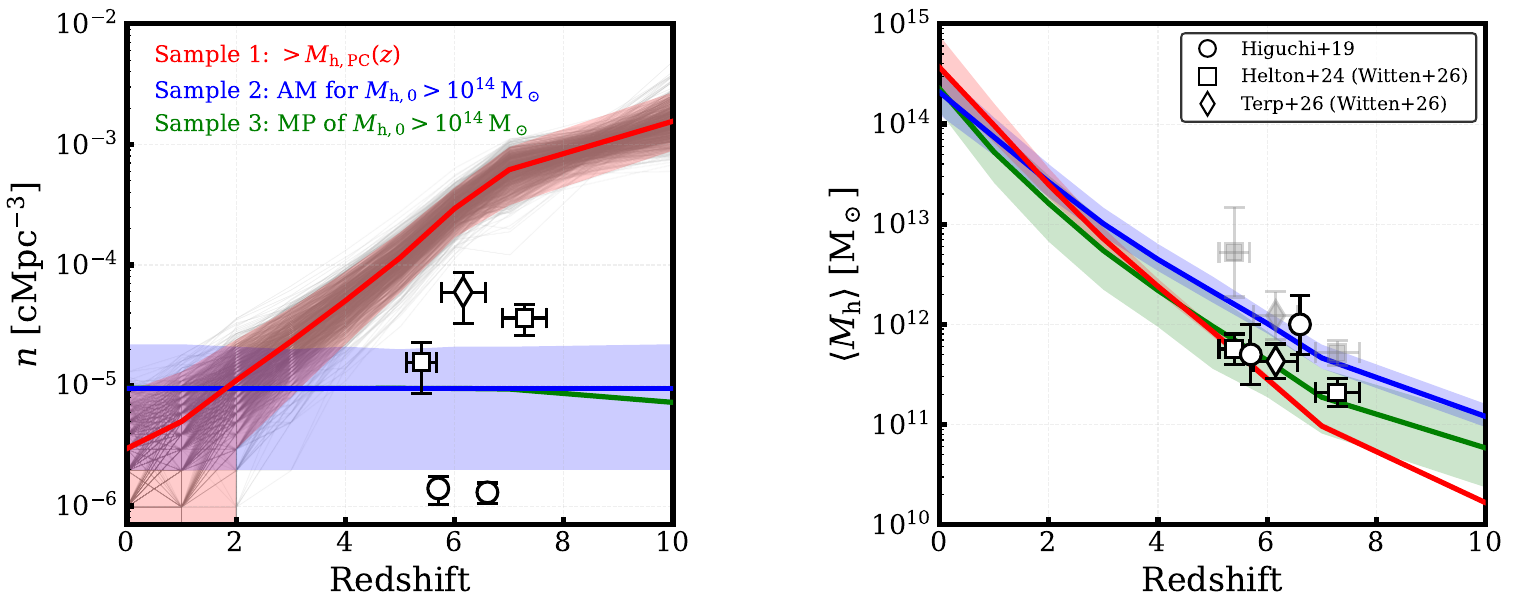}
\vspace{0.0cm}
\caption{Abundance (left) and total mass (right) of protocluster candidates. The solid lines show the number density (left) and total mass $M_{200}$ (right) of our three samples from the FLAMINGO simulation (L1\_m8) as a function of redshift: Sample 1 (mass-based; blue), Sample 2 (abundance-matched; orange), and Sample 3 (true main progenitor; green). The shaded bands represent the $2\sigma$ cosmic variance from 1,000 sub-volumes of $(100\,\mathrm{cMpc})^3$, with individual sub-volume number densities shown for Sample 1. Observational samples from \citet{Higuchi2019} (circles), \citet{Helton2024b} (square), and \citet{Terp2026} (diamonds) are overplotted. In the right panel, grey symbols show the original values, while white symbols show the corrected masses from \citet{Witten2026}. The observed abundance lies between Sample 1 and Sample 2, suggesting that current observational searches select both true cluster progenitors and contaminants. The corrected masses from \citet{Witten2026} bring the observations into agreement with our simulated samples, validating our sample selection.}
\label{fig1_selection}
\end{figure*}

Another recent systematic search was carried out by \citet{Terp2026} using rest-frame optical JWST/NIRCam grism spectroscopy of the Abell 2744 lensing field. Applying an FoF algorithm with a redshift-dependent linking length ranging between 5 and 8 cMpc, they identified six galaxy overdensities at $z\sim5.5-7$, five of which are robust systems with $N_{\mathrm{gal}}\geq20$ spanning $z=5.66$ to $6.77$, plus an additional structure at $z=7.88$. We only take the robust systems for our analysis. The survey covers an area of approximately $30\,\mathrm{arcmin}^2$, and the total effective comoving volume between redshift 5.5 and 7 is $8.4\times10^4\,\mathrm{cMpc}^3$. Total halo masses inferred from kinematics, UV luminosities, and stellar mass scaling consistently yield $M_{\mathrm{h}}\sim10^{12}-10^{14}\,\mathrm{M}_{\odot}$.

As pointed out and demonstrated by \citet{Lim2024}, there is a general issue of inconsistent apertures used to discover protoclusters and estimate their integrated physical properties, which challenges robust data analysis and comparisons with models. \citet{Lim2024} showed that the aperture sizes adopted among studies range from the virial radius of the core halo (a few tens of pkpc) to the entire Lagrangian region that will eventually collapse into a cluster (of order 10 cMpc), spanning nearly two orders of magnitude. They further demonstrated that this range of aperture sizes can lead to order-of-magnitude differences in estimates of total mass, stellar mass, and SFR for a protocluster candidate at $z\gtrsim4$. These findings were independently confirmed by \citet{Witten2026}, who corrected the mass estimates of \citet{Helton2024b} and \citet{Terp2026} by using simulation predictions of the stellar-to-halo mass relation (SHMR) and assigning to each protocluster the total mass associated with its most massive member galaxy. The corrected masses are smaller than the values reported in the original papers by 0.5 to 1.5 dex, as shown in Fig.~\ref{fig1_selection}. We use the original values from \citet{Higuchi2019} because they already accounted only for the most massive member galaxy. For the protocluster candidates from \citet{Higuchi2019} and \citet{Helton2024b}, we present the mean and standard deviation for each of the two redshift bins, while all five samples from \citet{Terp2026} are combined and represented by a single data point.


\section[Methods]{Methods}
\label{sec_method}

\subsection{Definition of FLAMINGO protocluster samples}
\label{ssec_sample}

There are several possible definitions of protoclusters that are adopted in the literature to allow comparison between models and observations. While this ambiguity in definition provides some fundamental challenges in protocluster studies \citep[see e.g.,][for detailed discussions]{Lim2021, Lim2024}, it offers, at the same time, a range of uncertainties that is useful and one should keep in mind when making comparisons with data.

In this study, we explore three definitions to select simulated protocluster samples from FLAMINGO, which cover theoretical, observational, and empirical ways to define and identify protoclusters from observations and models. First, we construct `Sample 1', our fiducial sample, by selecting all haloes in each redshift snapshot that meet the mass criteria. Specifically, we select FLAMINGO haloes of $M_{\rm h} = (3-10)\times 10^{14}\,{\rm M}_\odot$ (Virgo-like) at $z=0$. We then follow their merger trees to identify their main progenitors at high redshifts and compute their mean mass. As a final step, at each redshift, we select all isolated haloes above the mean mass minus 0.3 dex (as a lower limit to account for scatter), $M_{\rm h, PC}(z)$, and define them as protoclusters (Sample 1). This selection, which largely follows \citet{Chiang2013} (see also \citealt{Lim2024}), approximates the most common way to identify protocluster candidates from observations and predict their fate by comparing with models \citep[e.g.,][]{Laporte2022, Helton2024b}. We therefore consider Sample 1 as our fiducial selection of FLAMINGO protoclusters.

Sample 2 is constructed via abundance matching: we define and count as clusters all haloes at $z=0$ with $M_{\rm h} > 10^{14}\,{\rm M}_\odot$. Their number count is 9,485 in the FLAMINGO L1\_m8 simulation. At higher redshifts, we sort all haloes by mass and select the 9,485 most massive haloes as Sample 2. This selection strictly reflects a common assumption in protocluster studies that massive structures at high redshift are likely progenitors of those at low redshift, i.e., no significant alteration in the rank and scale of systems through their evolutionary track.

Finally, Sample 3 consists solely of the main progenitors of clusters, i.e., all haloes at $z=0$ with $M_{\rm h} > 10^{14}\,{\rm M}_\odot$. This selection may be viewed as the closest to the theoretical meaning of protoclusters: progenitors of clusters.

Note that all three selections are solely based on total mass and do not depend on any galaxy properties. This is by design to minimize uncertainties from baryonic processes, for which our understanding is far from complete, particularly at high redshifts. The goal of this study is to investigate the evolution, including mergers, of high-$z$ massive systems and their implications for observations and their abundance at later stages. We focus our analysis on dark matter haloes, whose properties we believe to be accurately modelled, including their abundance, clustering, and evolution over time.

\subsection{Evaluation of cosmic variance}
\label{ssec_CV}

We also evaluate the impact of cosmic variance on the evolution of protoclusters. This is particularly important because most high-redshift protocluster candidates from observations were identified from relatively small search volumes of order $\sim10^6\,{\rm cMpc}^3$. We thus divide the L1\_m8 box into 1,000 sub-volumes of $(100\,\mathrm{cMpc})^3$ and compute the mean properties within each sub-volume as well as the scatter among them to assess cosmic variance.

\section[Results]{Results}
\label{sec_results}

\subsection{Abundance and total mass of protoclusters}
\label{ssec_abun}

The abundance and total mass of protoclusters are two of the most basic properties analysed and compared across studies. Some high-redshift protocluster candidates, in particular, have been found to be in tension even with the simple theoretical halo mass function, as noted by \citet{Lim2024} and \citet{Witten2026}. \citet{Lim2024} demonstrated that the discrepancy cannot be explained by cosmic variance, reaching up to $3\sigma$. Both \citet{Lim2024} and \citet{Witten2026} revealed that using consistent apertures, combined with accounting for scatter in evolutionary paths, can significantly mitigate the tension.

The situation is well demonstrated in Fig.~\ref{fig1_selection}, where the left and right panels compare the number density and total mass, respectively, between simulated and observed protoclusters. The results from the individual sub-volumes (for Sample 1 only), as well as the $2\sigma$ cosmic variance among the sub-volumes, are presented alongside the quantities from the entire simulation box. In the left panel, the abundance of protocluster candidates from \citet{Helton2024b} and \citet{Terp2026} is 1 to $4\sigma$ above that of Sample 2, which is essentially the number density of clusters at redshift 0.\footnote{The number count of Sample 3 is by design identical to Sample 2, except at $z>7$ where about 10 per cent of their progenitors fall below the numerical resolution and are thus not tracked by the merger tree.} On the other hand, their number counts are significantly below those of Sample 1. As further discussed in Sect.~\ref{sec_discussion}, this is presumably because observations with sufficiently large search areas, including the three studies considered here, identify a region of multiple galaxies as a single candidate. The abundance from the two observational studies lying between Sample 1 and Sample 2 implies that the methodologies employed by those observations may select massive objects that contain some protocluster candidates but also significant contamination from systems that will not become clusters at later times. In contrast, the systems from \citet{Higuchi2019} show a number count substantially below even that of $z=0$ clusters. While the origin of this lower abundance is unclear, based on the mass comparison in the right panel, \citet{Higuchi2019} may have identified more massive systems (at their redshifts) relative to the other two studies, which are rarer. The right panel of Fig.~\ref{fig1_selection} also indicates that our three definitions of simulated protocluster samples are broadly consistent in total mass with the observational candidates, but only after the corrections applied by \citet{Witten2026}. All these findings ensure that the range of our sample selections covers the observations broadly well, making them suitable for comparison and analysis.

It is worth noting, in the left panel of Fig.~\ref{fig1_selection}, that the rank order of the sub-volumes in terms of number density barely changes from $z\simeq7$ to $z\simeq3$, but varies significantly at $z\gtrsim7$. Although we confirmed the same trends for Samples 2 and 3 (not shown), this is an interesting trend for Sample 1 in particular, because Sample 1 does not track the same objects nor only the most massive systems. The rarity, and thus the scale, of the systems that Sample 1 traces across time also evolves significantly, as clearly seen in their number density. This indicates that objects between redshift 7 and 3 evolve coherently in both scale and time. However, their earlier and later evolution is substantially affected by scatter in their accretion history. At lower redshifts ($z\lesssim2$), the increased statistical noise largely explains the behaviour as the number density drops. However, as will be seen later, mergers with massive systems peak at $z\simeq1-2$, further altering the most massive objects across the sub-volumes. At very early times ($z\gtrsim7$), haloes grow rapidly even through smaller mergers and accretion, which introduces significant changes in their evolutionary tracks over a short time of $\sim300$ Myr. This re-emphasises that one should be careful when interpreting high-$z$ protoclusters and massive objects in general, particularly when attempting to connect them to their higher-redshift progenitors and later descendants.

\begin{figure}
\includegraphics[width=1.02\linewidth]{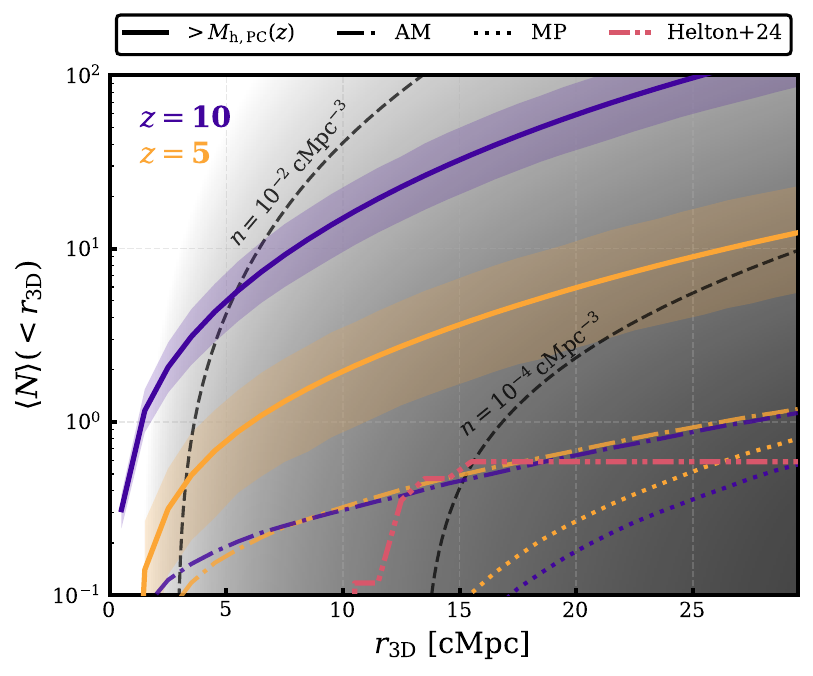}
\vspace{-0.3cm}
\caption{Average number of neighboring protocluster candidates, $\langle N \rangle$, within a 3D comoving radius $r_{\rm 3D}$, shown for our three FLAMINGO samples---mass-based: Sample 1 (solid), abundance-matched: Sample 2 (dashed), true progenitor: Sample 3 (dotted)---and the observational sample of \citet{Helton2024b} (dash-dotted). The shaded bands indicate $2\sigma$ cosmic variance for Sample 1, obtained from 1,000 sub-volumes of $(100\,\mathrm{cMpc})^3$. The black dashed lines and colour gradient show the volume density as a function of $r_{\rm 3D}$ and $\langle N \rangle$. Sample 1, our fiducial observationally motivated selection, has on average 2--10 neighbours within the typical protocluster scale of 10 cMpc. This strong clustering implies that a substantial fraction of observed candidates may later merge into more massive systems rather than surviving as isolated cluster progenitors. In contrast, the Helton et al. sample shows significantly weaker clustering, a discrepancy that cannot be explained by cosmic variance.}
\label{fig2_Nneighbor}
\end{figure}

\begin{figure*}
\includegraphics[width=0.75\linewidth]{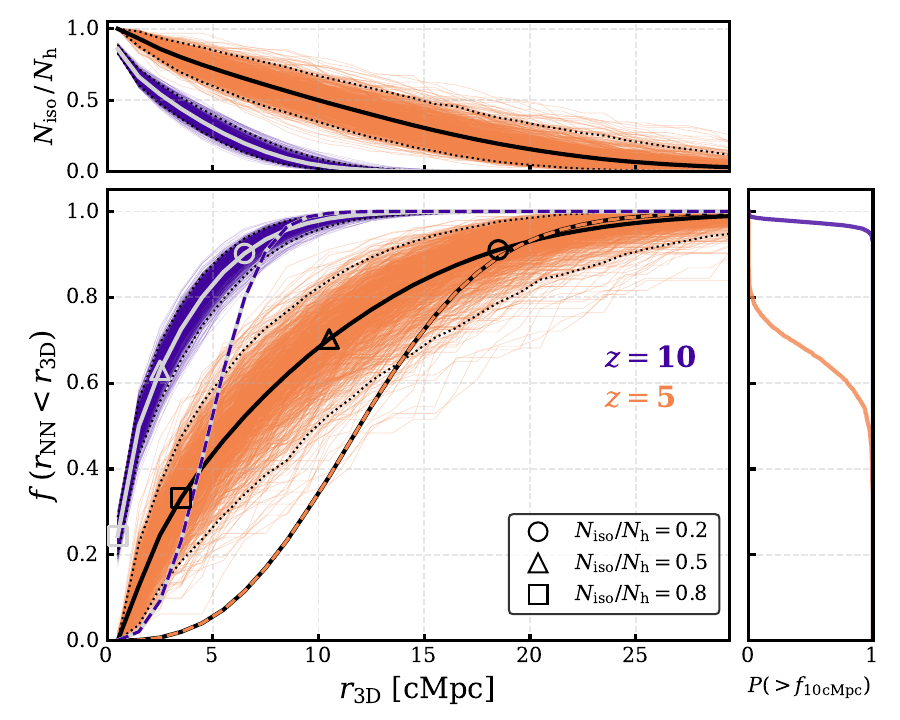}
\vspace{-0.0cm}
\caption{Cumulative fraction of protocluster candidates (main panel; for our fiducial sample, Sample 1, from the FLAMINGO simulation L1\_m8) as a function of the distance to their nearest neighbour, $r_{\rm NN}$, at $z=5$ and $z=10$. The thick lines show the mean for the whole simulation box, with the shaded bands representing $2\sigma$ cosmic variance from 1,000 sub-volumes of $(100\,\mathrm{cMpc})^3$; thin lines show individual sub-volume number densities. The dashed lines show the expectation for a random spatial distribution, obtained by randomly shuffling the 3D positions of the samples. The right panel shows the cumulative distribution across sub-volumes of the fraction of samples with $r_{\rm NN}\leq10$ cMpc. The top panel shows the fraction of systems identified as FoF-linked groups as a function of linking length $r_{\rm 3D}$. From the main panel, more than 90\% (70\%) of candidates at $z=10$ ($z=5$) have a massive neighbour within 10 cMpc, demonstrating that high-$z$ protoclusters are strongly clustered among themselves. This implies that a significant fraction of observational candidates may not be true cluster progenitors but will later merge into comparably or more massive systems.}
\label{fig25_NN}
\end{figure*}

\subsection{Clustering of protoclusters}
\label{ssec_clustering}

Here we investigate the clustering of protoclusters from both the simulations and observations. Although not yet widely recognised, clustering is potentially one of the most critical pieces of information for predicting the fate and evolution of protocluster candidates, as it relates to the possibility of later mergers with other massive systems nearby. Clustering and the possibility of late mergers also have direct links to the characteristics of observational candidate systems, including whether they are truly progenitors of clusters. If the chance of major mergers at later times is high, it likely means that those candidates are not systems that evolve into clusters and thus are not protoclusters by the strict theoretical definition. Later mergers may also explain the overabundance of observational candidates, as many of them may not survive as isolated systems as a result of mergers.

To this end, we first investigate the clustering of the protocluster samples, shown in Fig.~\ref{fig2_Nneighbor}. Specifically, the average number of neighbours within a 3D distance of each protocluster sample is shown for the three FLAMINGO samples and the observational sample of \citet{Helton2024b}. We exclude the other two observations in this comparison because \citet{Higuchi2019} samples may probe different (rarer) populations, for which we indeed find much lower clustering strength, and because \citet{Terp2026} contains too few systems. Figure~\ref{fig2_Nneighbor} shows that Sample 1, the most observationally motivated selection, has roughly between 2 (at $z=5$) and 10 (at $z=10$) neighbours on average within 10 cMpc, the typical protocluster size at these redshifts \citep{Chiang2013, Chiang2017}. Since the protocluster scale approximates the region that will collapse to form a single cluster by $z\simeq0$, our results indicate that at $z=5$, on average, two out of every three observational protoclusters (and about 90 per cent at $z=10$) will disappear over time via mergers into a nearby more massive candidate. Note also that the average number density (black dashed) of Sample 1 within the typical protocluster scale of 10 cMpc is between $10^{-3}$ and $10^{-2}$ cMpc$^{-3}$, higher than their global value (Fig.~\ref{fig1_selection}) by a factor of a few, meaning that high-$z$ protoclusters are clustered. Samples 2 and 3, which have lower global abundances by about an order of magnitude than Sample 1 at these redshifts, exhibit even stronger clustering, with the local number density within 10 cMpc being a few tens of times higher than the global value. However, because of their much lower global abundance, their merger chance is substantially lower than that of Sample 1 and almost negligible, implying that most of them will evolve as isolated systems through $z=0$ (regardless of whether they become clusters or lower-mass systems). This is particularly true for Sample 3, as expected, because they are selected as the true progenitors.

\begin{figure*}
\includegraphics[width=0.75\linewidth]{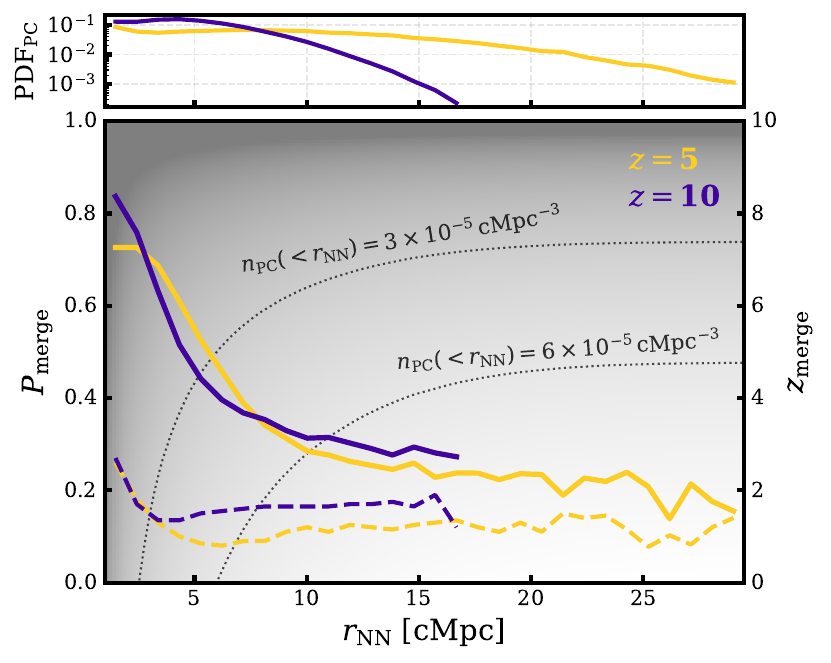}
\vspace{-0.0cm}
\caption{Merger probability and survival fraction as a function of nearest neighbour distance $r_{\rm NN}$ at $z=5$ and $z=10$, for our fiducial mass-based protocluster samples from the FLAMINGO simulation (L1\_m8). The main panel shows the probability that a protocluster candidate merges into a more massive system by $z=0$ ($P_{\rm merge}$; solid). The dashed lines (right axis) show the average redshift of the first such merger, $z_{\rm merge}$, which is typically $z\lesssim2$ and insensitive to $r_{\rm NN}$ except at very small separations. The dotted lines and colour gradient show the cumulative volume density of candidates that survive as isolated systems through $z=0$ as a function of $r_{\rm NN}$. The top panel shows the relative contribution to the number density of surviving (isolated) systems as a function of $r_{\rm NN}$. A candidate with a neighbour within 5 cMpc (10 cMpc) has a $\gtrsim50\%$ ($\gtrsim30\%$) chance of not surviving as an isolated cluster progenitor. Remarkably, even systems with very distant neighbours exhibit a floor of $\sim20\%$ merger probability, driven by later mergers with systems not yet massive at the selection epoch.}
\label{fig3_Pmerge}
\end{figure*}

For comparison, we also show the clustering of the observational sample of \citet{Helton2024b} in Fig.~\ref{fig2_Nneighbor}. Their sample consists of 17 protocluster candidates, thus providing insufficient statistics for a robust comparison. Nevertheless, their average number of neighbour candidates is significantly below the $2\sigma$ cosmic variance bands, essentially ruling out consistency with the model predictions. They have practically no neighbours within 10 cMpc (even though their chosen linking length is smaller by a factor of 2 to 3), meaning most of them are likely evolving as isolated systems. The data rather show a number of neighbours similar to that of Samples 2 and 3. The much lower global abundances of Samples 2 and 3 relative to the \citet{Helton2024b} systems (Fig.~\ref{fig1_selection}), however, indicate much weaker clustering for the observed protoclusters. This is surprising given that the selection by \citet{Helton2024b} is primarily based on the local galaxy overdensity with respect to the global mean, thus picking up the most clustered regions. While it is unclear whether this discrepancy arises from the selection or observational systematics, it suggests that clustering may provide additional information and constraints to complement global abundance for understanding protocluster candidates from observations.

Another way to probe the clustering of protoclusters is via the distance to their nearest neighbour, $r_{\rm NN}$, and via an FoF-like association. While the mean number of neighbours (as in Fig.~\ref{fig2_Nneighbor}) provides first-order clustering statistics, the distribution of $r_{\rm NN}$ and FoF groups offer higher-order information. It may also be more directly related to predicting the chance of mergers and evolution, under the assumption that if they merge, they will likely do so with their nearest neighbour.

Figure~\ref{fig25_NN} presents the cumulative fraction of simulated protoclusters in Sample 1 as a function of $r_{\rm NN}$. We analyse only Sample 1 here, as it is the selection that most resembles observational identifications of protoclusters and predictions of their fate. For each protocluster sample, we identify its nearest neighbour within Sample 1 and calculate the distance to it. We do this separately for each sub-volume to assess the impact of cosmic variance. We also repeat the experiment with randomly shuffled sample positions; the results are shown by dashed lines.

First, this confirms the prediction from Fig.~\ref{fig2_Nneighbor} that high-$z$ protoclusters are strongly clustered. They are not just clustered regions of galaxies individually, but are also clustered among themselves. This is clearly demonstrated by the much higher cumulative fraction at small $r_{\rm NN}$ relative to the random shuffle case. More than 90 per cent (70 per cent) of the samples have massive neighbours within 10 cMpc at redshift 10 (5). Given that the 10 cMpc scale typically assumed for protoclusters at these redshifts is the volume that later collapses to a single cluster, the results imply that a significant fraction of observational candidates may not be cluster progenitors but instead later merge into comparably or more massive systems. Cosmic variance cannot resolve this discrepancy: at $z=5$, more than 90 per cent of the sub-volumes have at least half of their samples with a neighbour within 10 cMpc (right panel).

In the top panel of Fig.~\ref{fig25_NN}, we show the fraction of systems identified as FoF-linked groups as a function of linking length $r_{\rm 3D}$. This is not exactly the same as one minus the cumulative fraction calculated above, because multiple haloes can be linked to the same FoF structure. The results show that, on average, about half (90 per cent) of protocluster candidates at $z\simeq5$ ($z\simeq10$) are `to-be' satellites embedded in more massive structures.

\subsection{Mergers of protocluster candidates}
\label{ssec_merger}

As discussed in Sect.~\ref{ssec_clustering}, clustering and mergers have direct impacts and implications for the characteristics and evolution of protocluster candidates from observations. Given the high expected number of neighbours shown in Sect.~\ref{ssec_clustering}, mergers indeed have significant impacts on these systems at later evolutionary stages. In this subsection, we directly investigate the chance and frequency of mergers and discuss their implications. We present results only for the fiducial sample, Sample 1.

For each protocluster, we trace its merger tree to see whether it survives as an isolated system through $z=0$ or is integrated into a more massive system and becomes a satellite. Figure~\ref{fig3_Pmerge} presents the fraction of simulated protocluster samples that later merge into a bigger system as a function of the distance to their nearest neighbour, $r_{\rm NN}$. Investigating mergers as a function of $r_{\rm NN}$ is motivated by the reasonable assumption that if a protocluster ever experiences a major merger(s) at a later time, it will most likely merge with its closest neighbour first. The results indicate that a high-$z$ protocluster candidate with its nearest neighbour within 5 cMpc (10 cMpc) has a $\gtrsim50\%$ ($\gtrsim30\%$) probability of not surviving as an isolated system, and thus not being a true cluster progenitor. Remarkably, the curve flattens at large $r_{\rm NN}$ and converges to about 20 per cent of samples always being absorbed into a bigger halo, regardless of their distance to neighbours. As will be seen shortly, this is due to mergers with nearby systems whose masses fall below the selection threshold at high redshift but grow rapidly to become more massive by the time of the mergers.

A $\gtrsim30$ per cent merger probability with bigger systems at $r_{\rm NN}=10$ cMpc does not mean that the abundance of protoclusters is overestimated by the same fraction, because the number of samples also varies with $r_{\rm NN}$. To account for this, the top panel shows the relative contribution to the number density of simulated protocluster samples that survive through $z=0$ as isolated systems at different $r_{\rm NN}$. This is essentially the fraction of samples in each $r_{\rm NN}$ bin multiplied by the merger probability for that bin. The cumulative number density of surviving protocluster samples is also shown as dotted lines in the main panel. Both results indicate that true (isolated through $z=0$) protocluster systems in the simulation have a fairly uniform contribution from those with $r_{\rm NN}\lesssim10$ cMpc to their final abundance. Their number density, however, gradually decreases and becomes almost negligible at $r_{\rm NN}\gtrsim30$ cMpc ($\gtrsim15$ cMpc) at redshift 5 (10), where the relative abundance is more than two orders of magnitude smaller than its peak value.

We also show, on the right-hand axis and by dashed lines, the average redshift of the first merger into a larger halo, $z_{\rm merge}$, at a given $r_{\rm NN}$. Mergers typically occur at relatively late times ($z<2$). Notably, $z_{\rm merge}$ is insensitive to the distance to the closest companion except when the neighbour is within 3 cMpc.

\begin{figure}
\includegraphics[width=1.0\linewidth]{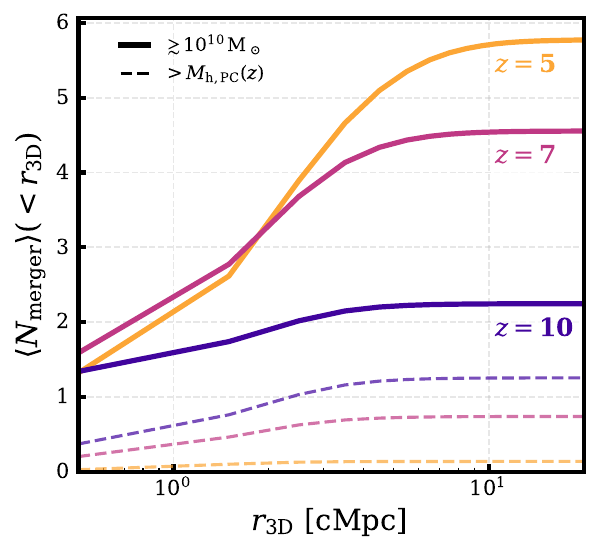}
\vspace{-0.3cm}
\caption{Cumulative number of later mergers into more massive systems as a function of the 3D separation at the selection redshift $z=5$, $z=7$, and $z=10$, for our fiducial mass-based protocluster sample from the FLAMINGO simulation (L1\_m8). The dashed lines count only mergers with systems that are already above the protocluster mass threshold $M_{\rm h,PC}(z)$ at the selection epoch, while the solid lines include all mergers regardless of the current mass of the merging partner. When considering only already-massive neighbours, candidates experience at most one merger on average. However, when including all mergers, a typical candidate at $z\simeq5$ ($z\simeq10$) undergoes $\sim6$ ($\sim2$) mergers into larger systems, indicating that later major mergers are dominated by systems not yet assembled as massive structures at the selection epoch. The merger count converges at separations $\gtrsim10$ cMpc, revealing a characteristic scale for structure formation beyond which mergers can be safely neglected.}
\label{fig4_Nmerger}
\end{figure}

Having examined the probability of mergers and surviving protoclusters, another key question regarding their later evolution is how many mergers each system will experience before reaching $z=0$. This has a more direct implication for the abundance of protocluster candidates. In Fig.~\ref{fig4_Nmerger}, we explore the cumulative number of later mergers as a function of 3D separation at the redshift of selection. We count only mergers with more massive haloes. Furthermore, the dashed lines count only mergers into systems whose mass is above $M_{\rm h,PC}(z)$ at the selection epoch, while the solid lines include all such mergers regardless of the current mass of the merging partner. The latter is not affected by numerical resolution because we only consider mergers into systems more massive than the simulated sample, which is already above the resolution limit and (almost) always grows in mass.

First, simulated protocluster samples mimicking the mass-based observational selection experience, on average, one merger or less with more massive systems among themselves at later times. This agrees with the finding of \citet{Terp2026} that their protocluster samples do not overlap within proximity. The merger count also decreases with decreasing redshift at which the samples are identified, with almost no mergers at $z\lesssim5$. When all mergers are considered regardless of the current mass of the merging partner, however, FLAMINGO predicts that a typical protocluster candidate at $z\simeq5$ ($z\simeq10$) undergoes about 6 (2) mergers into larger systems. This indicates that later major mergers of protoclusters are dominated by systems that are not yet assembled and identified as massive at the selection redshift. Again, this highlights the fundamental uncertainty in selecting high-redshift objects and projecting their evolution based on the mean accretion history that ignores scatter.

Second, it is interesting to note that, in all cases and at all selection redshifts, the cumulative number of mergers converges at separations $\gtrsim10$ cMpc, i.e., there are practically no mergers with massive objects from beyond the protocluster scale typically assumed at these redshifts. This holds even though Sample 1 is not a selection of true protoclusters but is solely based on mass, meaning that many of them do not evolve into clusters but into smaller systems. These results indicate that there is a characteristic scale of $\sim10$ cMpc that governs the cosmic evolution of structures over a wide range of masses. This has implications for observations with systematic volume searches for protoclusters: whether their candidates are true cluster progenitors or not, the possibility of mergers between them can be safely neglected as long as they are separated by more than $\sim10$ cMpc. Of course, a key difference is that massive structures like true protoclusters `clear out' the surrounding volume on this scale by accreting other systems, whereas the samples studied here are not the most massive in their volume and go through multiple mergers to later become satellites.

\section[Discussion]{Discussion}
\label{sec_discussion}

\subsection{Are observational protocluster candidates at $z\gtrsim5$ really progenitors of clusters?}
\label{ssec_what}

Recently, a question has been raised as to whether protocluster candidates discovered by observations are true progenitors of clusters. Although the mismatch between theoretical and observational definitions of protoclusters has been a long-standing fundamental issue, recent studies extending to extremely high redshifts ($z\gtrsim5$, even up to 10) introduce even more uncertainties. Given the time span across this redshift range, many physical processes---such as the early formation of galaxies and the evolution of their properties, as well as the evolution of large-scale structures and scatter in assembly history, where our current understanding is limited---may have significant impacts on the interpretation of these studies.

Partial answers to this question have been suggested by \citet{Lim2024}, \citet{Morokuma-Matsui2025}, and \citet{Witten2026}, and also by \citet{Remus2023} for slightly lower redshifts. Most of these studies share common concerns, casting doubt on whether a significant fraction of the protocluster candidates identified at $z\gtrsim5$ with current observational methodology are true cluster progenitors. Observations usually determine the fate of their samples---whether they will later evolve into cluster-scale systems or not---based on mass, even though their initial identification as candidates is often made through galaxy overdensities. Furthermore, \citet{Lim2024} (and also \citet{Witten2026}) pointed out an inconsistency commonly found in observational procedures: a variety of apertures are adopted but not properly corrected for when comparing with model predictions. Additionally, \citet{Witten2026} nicely demonstrated that true cluster progenitors have masses close to the observational candidate systems, but they constitute only a small fraction of all objects of similar mass. \citet{Witten2026} showed that a majority of high-$z$ systems with masses similar to the observational candidates evolve into groups (with total masses between $10^{13}$ and $10^{14}\,{\rm M}_\odot$) by $z\simeq0$ instead. This essentially means that any selection or interpretation primarily based on mass, as in most high-$z$ studies, will inevitably fail because there is too much contamination in a given mass bin at high-$z$ due to scatter in later evolution. This leads to the second part of the discussion: what is the optimal way to select true high-$z$ progenitors of clusters?

\subsection{What is an optimal way to identify true cluster progenitors at high-$z$?}
\label{ssec_optimal}

One possible addition to mass is clustering. Indeed, many observational studies first identify their candidates primarily based on the local galaxy overdensity, $\delta_{\rm gal}$. As demonstrated in studies such as \citet{Chiang2013}, \citet{Lovell2018}, and \citet{Morokuma-Matsui2025}, however, there is significant scatter in $\delta_{\rm gal}$ (over $10-15$ cMpc) for true cluster progenitors at high-$z$. This is also well reflected among observational studies, which often present large scatters in the galaxy overdensity of candidates even within a single systematic search. Even worse, similar to the situation with mass, even regions of high $\delta_{\rm gal}$ have substantial contamination, with many non-cluster volumes often possessing high galaxy clustering. For instance, there is only about a 50 per cent chance that a region with $\delta_{\rm gal}$ as high as 8 at redshift 5 evolves into a cluster \citep{Chiang2013}. Predicting the exact scale to which a system evolves---e.g., a Coma-like cluster, a Virgo-like cluster, or a group---is even more uncertain.

Relying on $\delta_{\rm gal}$ is also not well justified by the conventional definition of a cluster (and thus a protocluster), which depends solely on the total virialised mass of a system. Depending on the evolutionary stage of a protocluster, which we do not understand well, a region may have most of its galaxies already merged into a few galaxies \citep[e.g.,][]{Muldrew2015}, in which case the system might not be accurately identified via $\delta_{\rm gal}$. In many cases, the error in $\delta_{\rm gal}$ is also significant because galaxies at high redshift are faint, with a majority of the population undetected due to observational limits. This results in the identification of protoclusters from only a few or even a single galaxy, as indeed often found in observational studies. $\delta_{\rm gal}$ for a given protocluster region also depends strongly on the type of tracers used \citep[e.g.,][]{Chiang2013}. Furthermore, \citet{Baxter2025} pointed out that the peak in galaxy density is typically offset from the peak in mass distribution by more than 3 to 10 cMpc. Finally, identifying protoclusters requires a sufficiently large search area to cover the protocluster scale of about 10 cMpc in order to detect member galaxies and calculate $\delta_{\rm gal}$ properly, which is often lacking in current high-$z$ observations \citep[see also][]{Baxter2025}.

While both mass and galaxy overdensity---the two most commonly used identification methods---are problematic, some studies such as \citet{Chiang2013} and \citet{Remus2023} have suggested that the total mass enclosed within roughly a protocluster scale is a much more reliable predictor of a system's fate (see also \citealt{Witten2026}). This is consistent with the natural prediction that the mass of the final system is determined by the total mass enclosed within the volume that eventually collapses into the bound object. The requirement for this is, of course, a sufficiently large survey volume to observe all potential member galaxies. It is also necessary to probe down to the faintest galaxies because, due to their higher number density, their contribution to integrated properties (including total mass) within a volume can still be significant \citep[e.g.,][]{Muldrew2015, Lim2024}.

In this regard, our results, e.g., from Fig.~\ref{fig4_Nmerger}, confirm that it is important to scan out to at least 10 cMpc, as many mergers occur later on smaller scales. This is consistent with the typical protocluster scale found by \citet{Chiang2013}, \citet{Chiang2017}, and \citet{Lovell2018}, although our selection is based primarily on mass, closer to empirical observational selections rather than true progenitors, thus providing a more general implication. An observation with a smaller search area would run a high risk that its candidate will merge with other undetected massive neighbours, leaving the prediction of its fate highly uncertain. Ideally, an observation should be able to identify all galaxies within $\sim10$ cMpc, sum their masses, and compare the total to the mass of $z\simeq0$ clusters. At a minimum, it must ensure that the central galaxy is the most massive one within 10 cMpc, i.e., satisfy an isolation criterion.

On the other hand, our results also indicate that it may not be necessary to increase the survey volume beyond $\sim10$ cMpc. Some studies have reported multiple protocluster candidates relatively close to each other (within a few tens of cMpc; e.g., \citealt{Higuchi2019} and \citealt{Helton2024b}), raising questions of whether those systems should be defined and counted individually or are part of the same structure that will later emerge as a single system through mergers. Based on the FLAMINGO predictions, however, mergers between systems more than 10 cMpc apart are very rare and can be safely neglected, not affecting protocluster evolution.

\section[Summary]{Summary and conclusions}
\label{sec_summary}

In this study, we have used the FLAMINGO hydrodynamical simulations to investigate the evolution and clustering properties of high-redshift protoclusters, comparing three different selection methods that span theoretical, observational, and empirical definitions. Our analysis focuses on redshifts $z \gtrsim 5$, where recent JWST and Subaru observations have identified numerous protocluster candidates, and we assess whether these systems are likely true progenitors of today's galaxy clusters.

Our key findings are as follows. First, the abundance of protocluster candidates from observations lies between our fiducial mass-selected sample and the abundance-matched sample, suggesting that current observational methodologies identify massive systems that include both true cluster progenitors and substantial contamination from structures that will not evolve into clusters by $z=0$. The total masses of observed candidates, after applying the aperture corrections advocated by \citet{Witten2026}, are broadly consistent with all three of our simulated samples, validating the range of our selections for comparison.

Second, we demonstrate that high-redshift protoclusters are strongly clustered. Our simulated samples show on average 2--10 neighbours within the typical protocluster scale of 10 comoving Mpc between $z=5$ and $10$. This implies that a significant fraction of observationally selected protocluster candidates may later merge into more massive systems rather than surviving as isolated cluster progenitors. In contrast, the observational sample of \citet{Helton2024b} exhibits significantly weaker clustering than our model predictions, a discrepancy that cannot be explained by cosmic variance alone and suggests that clustering may provide an additional, underutilized constraint for evaluating protocluster candidates.

Third, we quantify the merger probability as a function of the distance to the nearest neighbour. A protocluster candidate with a neighbour within 5 cMpc (10 cMpc) has a $\gtrsim 50\%$ ($\gtrsim 30\%$) probability of not surviving as an isolated system to $z=0$. Notably, even systems with very distant neighbours have a floor of about $20\%$ merger probability, driven by later mergers with systems that are not yet massive enough at the time of selection. The majority of these mergers occur at relatively late times ($z < 2$), and the cumulative number of mergers converges at separations $\gtrsim 10$ cMpc, indicating a characteristic scale governing structure formation across a wide range of masses.

Fourth, while typical protocluster candidates identified via mass-based selection experience on average one or fewer mergers with comparably massive systems at later times, when including mergers with all more massive haloes regardless of their current mass, the number increases to approximately 2--6 mergers depending on redshift. This highlights a fundamental uncertainty: later major mergers are dominated by systems that are not yet assembled as massive structures at the epoch of observation. Consequently, projecting the fate of high-redshift candidates based solely on their current mass or galaxy overdensity is subject to large scatter and potential systematic bias.

Looking forward, our results suggest that an optimal strategy for identifying true cluster progenitors at high redshift should combine mass estimates with isolation criteria within a radius of at least 10 comoving Mpc---the characteristic scale beyond which later mergers become negligible. Surveys that cannot probe to this scale or that do not account for undetected faint galaxies risk significant contamination and uncertain evolutionary predictions. Future JWST wide-field surveys and the next generation of 30-metre class telescopes, capable of mapping galaxy distributions over larger volumes and to fainter limits, will be essential to robustly connect high-redshift protocluster candidates to their eventual fate as clusters or groups in the local Universe.

\section*{ACKNOWLEDGEMENTS}

SL acknowledges support by the Science and Technology Facilities Council (STFC) and by the UKRI Frontier Research grant RISEandFALL. We thank Callum Witten for insightful discussions. This work used the DiRAC@Durham facility managed by the Institute for Computational Cosmology on behalf of the STFC DiRAC HPC Facility (\url{www.dirac.ac.uk}). The equipment was funded by BEIS capital funding via STFC capital grants ST/K00042X/1, ST/P002293/1, ST/R002371/1 and ST/S002502/1, Durham University and STFC operations grant ST/R000832/1. DiRAC is part of the National e-Infrastructure. This work is partly funded by research programme Athena 184.034.002 from the Dutch Research Council (NWO). We acknowledge the Virgo Consortium for making their simulation data available. The FLAMINGO simulations were performed using the Durham Memory Intensive system managed by the Institute for Computational Cosmology on behalf of the STFC DiRAC facility (www.dirac.ac.uk).

\section*{DATA AVAILABILITY}

The data underlying this article will be shared on reasonable request to the corresponding author.

\bibliographystyle{mnras}
\bibliography{GalEnv.bib}

\appendix


\label{lastpage}

\end{document}